\documentclass[apj]{aastex631}
\usepackage{amsmath}

\shortauthors{LAPI ET AL.}
\shorttitle{NON-UNIVERSAL AND NON-STATIONARY HALO MASS FUNCTION}

\begin{document}

\title{A Stochastic Theory of the Hierarchical Clustering. III. \\ The Non-universality and Non-stationarity of the Halo Mass Function}

\author[0000-0002-4882-1735]{Andrea Lapi}
\affiliation{SISSA, Via Bonomea 265, I-34136 Trieste, Italy}\affiliation{IFPU - Institute for fundamental physics of the Universe, Via Beirut 2, 34014 Trieste, Italy}\affiliation{INFN-Sezione di Trieste, via Valerio 2, 34127 Trieste,  Italy}\affiliation{INAF/IRA, Istituto di Radioastronomia, Via Piero Gobetti 101, 40129 Bologna, Italy}

\author[0000-0002-3515-6801]{Tommaso Ronconi}
\affiliation{SISSA, Via Bonomea 265, I-34136 Trieste, Italy}\affiliation{IFPU - Institute for fundamental physics of the Universe, Via Beirut 2, 34014 Trieste, Italy}

\author[0000-0003-1186-8430]{Luigi Danese}
\affiliation{SISSA, Via Bonomea 265, I-34136 Trieste, Italy}\affiliation{IFPU - Institute for fundamental physics of the Universe, Via Beirut 2, 34014 Trieste, Italy}

\begin{abstract}
In the framework of the stochastic theory for hierarchical clustering, we investigate the time-dependent solutions of the Fokker-Planck equation describing the statistics of dark matter halos, and discuss the typical timescales needed for these to converge toward stationary states, far away enough from initial conditions. Although we show that the stationary solutions can reproduce the outcomes of state-of-the-art $N-$body simulations at $z\approx 0$ to a great accuracy, one needs to go beyond to fully account for the cosmic evolution of the simulated halo mass function toward high-redshift. Specifically, we demonstrate that the time-dependent solutions of the Fokker-Planck equation can describe, for reasonable initial conditions, the non-universal evolution of the simulated halo mass functions. Compared to standard theoretical estimates, our stochastic theory predicts a halo number density higher by factor of several toward $z\gtrsim 10$, an outcome which can be helpful in elucidating early and upcoming data from JWST. Finally, we point out the relevance of our approach in designing, interpreting and emulating present and future $N-$body experiments.
\end{abstract}

\keywords{Cosmology (343) --- Dark matter (353)}

\section{Introduction}\label{sec|intro}

According to the standard cosmological framework, dark matter (DM) halos are thought to originate by the collapse of patches from an initial, closely Gaussian perturbation field. However, as demonstrated by many extensive $N-$body simulations (e.g., see textbooks by Mo et al. 2010 and Cimatti et al. 2020; also comprehensive reviews by Vogelsberger et al. 2020, Angulo \& Hahn 2022 and references therein), the detailed evolution of perturbations
and (proto)halos ultimately depends on a variety of effects. First, the role of initial conditions is crucial, in that a perturbation is more prone to collapse if it resides within a sufficiently over-dense region of the initial density field. This was actually the foundational idea of pioneering estimates for the halo abundance (see Press \& Schechter 1974), subsequently refined in terms of the excursion set approach (see Bond et al. 1991; Lacey \& Cole 1993; Mo \& White 1996) to avoid the double counting of over-dense regions overlapped with, or embedded within, larger collapsing ones (the so-called cloud-in-cloud issue).

In addition, the shape of (proto)halos may be also relevant, in that they tend to be ellipsoidal in shape (and especially so for smaller halos collapsing at late times), and this may influence the collapse efficiency and timescales (see Sheth \& Tormen 1999, 2002). Moreover, collapse locations may be special points in the initial perturbation field, such as peaks in density or in energy (e.g., Bardeen et al. 1986; Paranjape \& Sheth 2012; Lapi \& Danese 2014; Musso \& Sheth 2021). Finally, other nonlinear effects may influence the collapse of perturbations, such as mergers, tidal forces, dynamical friction, relaxation, local environment, angular momentum acquisition and dissipation, velocity fields, clumpiness, etc.; these involve different spatial/temporal scales, and enforce a considerable variance in the collapse of different (proto)halo patches in the Universe.

Thus it should appear evident that the collapse and evolution of DM halos constitute an inherently stochastic process, originating not only by a degree of randomness in the initial conditions, but also by the complexity of deterministic processes affecting the ensuing collapse. As a consequence, the fine details of the evolution for individual halos at different spatial locations and cosmic times are, for all practical purposes, difficult to follow and/or model ab initio in (semi-)analytic terms.
In this vein, Lapi \& Danese (2020, 2021; hereafter LD20) have submitted that, if one is mainly interested in the statistical properties of the halo population as a whole, an effective description of (proto)halo collapse can
be conveniently provided via a mean-field stochastic theory. Specifically, the average macroscopic dynamics of a (proto)halo mass $M(t)$ can be effectively described in terms of fluctuations driven by an appropriate noise term; the latter yields an average drift of the masses toward larger values, which renders the expected hierarchical clustering behavior of the halo population.

The situation is somewhat analogous to the classic description of Brownian motion: a microscopic particle immersed in a fluid continuously undergoes collisions with the fluid molecules; the resulting motion, despite being deterministic, appears to be random at the macroscopic level, especially to an external observer who has no access to the exact positions and velocities of the innumerable fluid molecules and to the initial conditions of the particle. In the way of a statistical description, the problem is effectively treated via a stochastic differential equation driven by a fluctuating
white noise, which allows to implicitly account for the complex microscopic dynamics of the system. Note that often the system’s state influences the intensity of the driving noise, like when the Brownian fluctuations of a microscopic particle near a wall are reduced by hydrodynamic interactions, so that the noise becomes multiplicative in terms of a non-uniform diffusion coefficient. Similar stochastic models with multiplicative noise have been employed to describe a wide range of physical phenomena, from Brownian motion in inhomogeneous media or in close approach to physical barriers, to thermal fluctuations in electronic circuits, to the evolution of stock prices, to computer science, to the heterogeneous response of biological systems and randomness in gene expression (e.g., Risken 1996; Reed \& Jorgensen
2004; Mitzenmacher 2004; Paul \& Baschnagel 2013).
Note that in many contexts the validity of the stochastic equations designed on purpose to describe the macroscopic dynamics is checked by comparison with observations and/or numerical simulations. In the cosmological evolution of halos, $N-$body simulations provide the most relevant testbeds, although any comparison must be performed with care since their outcomes are dependent on many subtle details such as the limited number of simulated particles (i.e., mass resolution), the treatment of gravity on small scales (i.e., softening length), the sample statistics (i.e., simulated volumes), the background cosmology, etc.

In LD20 we have shown a preliminary comparison of the halo mass functions predicted by our stochastic theory with the results of a few classic $N-$body simulations, including Sheth \& Tormen (1999), Bhattacharya et al. (2011), and Watson et al. (2013). The net outcome is that appropriately gauging the parameters which regulate the mass and redshift dependence of the multiplicative noise term, our approach is capable of reproducing the simulation results, at least at $z\approx 0$.  So far our analysis has been performed by only exploiting the stationary, steady-state solutions of the Fokker-Planck equation describing the stochastic dynamics;
however, a detailed comparison with the most recent $N-$body simulations (e.g., Shirasaki et al. 2021; Ishiyama et al. 2021), which probe extended redshift and mass ranges, shows that one needs to go beyond.
In particular, recent simulations have rekindled the interest and deepened the focus on the non-universality of the halo mass function; this means that the latter cannot be described in terms of an universal function $f(\nu)$ of a variable $\nu$ incorporating all its redshift and mass dependencies.
To interpret non-universality has proven a difficult task for standard approaches to the halo statistics, like the excursion set approach.

The main aim of the present work is to show that the non-universal behavior measured in simulations is naturally expected and quantitatively reproduced by our stochastic framework; specifically, it emerges whenever the solutions of the Fokker-Planck equation describing the stochastic dynamics have not yet converged to a stationary state, as it occurs for reasonable initial conditions in the standard cold DM cosmological framework.
The plan of the paper is straightforward: in Sect. \ref{sec|theory} we recall the basic formalism of our stochastic theory, focusing on non-stationary solutions of the Fokker-Planck equation and on the associated halo mass functions; in Sect. \ref{sec|Nbody} we quantitatively compare stationary and non-stationary mass functions, and highlight that the latter can reproduce, for reasonable initial conditions, the non-universal behavior as measured in $N-$body simulations;  in Sect. \ref{sec|summary} we discuss and summarize our main findings. In the Appendix we present an approximate expression for the non-stationary solutions of the Fokker-Planck equation with space-dependent drift, which is exploited in the main text. Throughout this work, we adopt the standard flat $\Lambda$CDM cosmology (Planck Collaboration 2020) with rounded parameter values: matter density $\Omega_{\rm M}=0.3$, dark energy density $\Omega_{\Lambda}=0.7$, baryon density $\Omega_{\rm b}=0.05$, Hubble constant $H_0=100\, h$ km\ s$^{-1}$ Mpc$^{-1}$ with $h=0.7$, and mass variance $\sigma_8=0.8$ on a scale of $8\, h^{-1}$ Mpc.

\newpage

\section{Stochastic theory of hierarchical clustering}\label{sec|theory}

In this section we recall the basics of the stochastic theory for hierarchical clustering developed in LD20; specifically, here we aim to highlight the intrinsic non-stationary behavior in the solutions of the Fokker-Planck equations, which regulate the halo mass function and its mass and redshift dependencies.

\subsection{Basic formalism}\label{sec|basics}

The proposal put forward by LD20 consists in describing the evolution of the halo population in terms of a stochastic formalism. In particular,
the mass $M(t)$ of a (proto)halo patch is promoted to a variable fluctuating along the cosmic time $t$, as ruled by the stochastic differential equation (in the Stratonovich convention)
\begin{equation}\label{eq|basic_mass}
\frac{\rm d}{{\rm d}t}\, M = \frac{\sigma^2}{|{\rm d}\sigma/{\rm d}M|}\, \frac{1}{\delta_c(t)}\, \left|\frac{\dot\delta_c(t)}{\delta_c(t)}\right|^{1/2}\, \eta(t)~,
\end{equation}
where $\eta(t)$ is a gaussian white noise with ensemble-average properties $\langle \eta(t)\rangle=0$ and  $\langle\eta(t)\eta(t')\rangle=2\, \delta_D(t-t')$.

In the above equation, the quantity $\delta_c(t)$ is the critical threshold for collapse, which takes on present values around $\delta_c(t_0)\approx 1.68$ and scales with cosmic time $t$ as $\delta_c(t)\propto D^{-1}(t)$ in terms of the growth factor $D(t)$ for linear perturbations.
In addition, $\sigma(M)$ is the mass variance filtered on the scale $M$, defined as
\begin{equation}\label{eq|variance}
\sigma^2(M) = \frac{1}{(2\pi)^3}\,\int{\rm d}^3k\, P(k)\,\tilde{W}_M^2(k)~,
\end{equation}
where $P(k)$ is the power spectrum of density fluctuation smoothed via a Fourier-transformed window function $\tilde{W}_M^2(k)$ whose volume in real space encloses the mass $M$; we note that any power spectrum or filter shapes can in principle be adopted in our framework.
For definiteness, and in order to comply with choices largely implemented in $N-$body simulations, we employ hereafter the classic cold DM power
spectrum by Bardeen et al. (1986). We also use the filter function proposed by Leo et al. (2018) with smooth shape $\tilde{W}_M^2(k) \propto [1+(k\,R_M)^{\omega_1}]^{-1}$, where the radius $R_M$ is simply related to the mass $M$ as $R_M\equiv (3\, M/4\,\pi\, \bar\rho)^{1/3}/\omega_2$ in terms of the average comoving matter density $\bar\rho$. For different parameters $(\omega_1,\omega_2)$ such a general shape can mimic the behavior of the filters usually adopted in the literature: top-hat in real space, $k-$sharp (i.e., top-hat in Fourier space), Gaussian; we adopt the fiducial parameters $(\omega_1,\omega_2)=(4.8,3.3)$ that are suggested by Leo et al. (2018) as the best choice for comparing with $N-$body mass functions. For the adopted power spectrum and filter shape, the relation $\sigma(M)$ from Eq.~(\ref{eq|variance}) is purely deterministic, featuring an inverse, convex, slowly-varying behavior.

Following LD20, the probability density $\mathcal{P}(M,t)$ for a region to enclose a mass between $M$ and $M+dM$ at time $t$ is derived by solving the Fokker-Planck equation associated to Eq.~(\ref{eq|basic_mass}), which reads (see Appendix A of LD20 for a primer on Fokker-Planck equations)
\begin{equation}\label{eq|FPmass}
\partial_t \mathcal{P}(M,t) = -\mathcal{T}^2(t)\, \partial_M \left[\mathcal{D}(M)\, \mathcal{D}'(M)\,\mathcal{P}(M,t)\right] +\mathcal{T}^2(t)\,\partial^2_M \left[\mathcal{D}^2(M)\, \mathcal{P}(M,t)]\right]~,
\end{equation}
where $\mathcal{D}(M)\equiv \sigma^2/|{\rm d}\sigma/{\rm d}M|$ and $\mathcal{T}(t)\equiv |\dot \delta_c|^{1/2}/\delta_c^{3/2}$. This must be supplemented by the natural boundary conditions $\mathcal{P}(\infty,t)=0$ and $P(M,t)=0$ for $M<0$, and by an initial condition $\mathcal{P}(M,t_{\rm in})=\delta_D(M-M_{\rm in})$ at a starting time $t_{\rm in}<t$;
the value $M_{\rm in}$ will be gauged against $N$-body simulations, as discussed in Sect. \ref{sec|Nbody}. Then the halo mass function is \emph{by definition} related to the solution $\mathcal{P}(M,t)$ of the Fokker-Planck equation via
\begin{equation}\label{eq|massfunc}
N(M,t) = \frac{\bar \rho}{M}\, \mathcal{P}(M,t)~.
\end{equation}

The solution of Eq.~(\ref{eq|FPmass}) can be obtained via a change of variables $X\equiv \int{\rm d}M/\mathcal{D}(M)=1/\sigma$, $Y\equiv \int{\rm d}t\, \mathcal{T}^2(t)=1/2\,\delta_c^2$ and $\mathcal{W}(X,Y)\equiv\mathcal{D}(M)\mathcal{P}(M,t)$, in such a way that it reduces to a standard diffusion equation $\partial_Y \mathcal{W}=\partial_X^2 \mathcal{W}$. Taking into account the aforementioned boundary conditions and coming back to the original variables (see LD20 for details) one easily finds the solution via a Fourier transform:
\begin{equation}\label{eq|sol_FPmass}
\mathcal {P}(M,t)=\frac{\delta_c}{\sigma^2}\,\left|\frac{{\rm d}\sigma}{{\rm d}M}\right|\, \frac{1}{\sqrt{2\pi\, (1-\delta_c^2/\delta_{c,\rm in}^2)}}\, \left\{\exp\left[-\frac{\delta^2_c}{2\sigma^2}\,\frac{(1-\sigma/\sigma_{\rm in})^2}{1-\delta_c^2/\delta_{c,\rm in}^2}\right]\, +\exp\left[-\frac{\delta^2_c}{2\sigma^2}\,\frac{(1+\sigma/\sigma_{\rm in})^2}{1-\delta_c^2/\delta_{c,\rm in}^2}\right]\right\}~,
\end{equation}
where $\delta_{\rm in}\equiv \delta_c(t_{\rm in})$ and $\sigma_{\rm in}=\sigma(M_{\rm in})$. Far away from the initial conditions $t\gg t_{\rm in}$ and $M\gg M_{\rm in}$ one has that $\sigma\ll \sigma_{\rm in}$ and $\delta_c\ll \delta_{c,\rm in}$, to yield the stationary solution\footnote{We stress that to attain stationarity the time derivative $\partial_t\mathcal{P}$ appearing on the left hand side of Eq.~(\ref{eq|FPmass}) must be negligible with respect to the terms on the right hand side. To a crude approximation, such a condition is met when the timescale $t_{\rm FP}\sim \mathcal{T}^{-2}(t)\propto \delta_c^3(t)/\dot\delta_c(t)<<t$ gets substantially smaller than the cosmic time $t$. In the standard $\Lambda$CDM cosmology the scaling $\delta_c\propto t^{-\zeta}$ with $\zeta\sim 2/3-1/2$ holds in the matter or dark energy dominated era, implying that stationarity may be attained only at sufficiently late times, typically far away from the initial $t_{\rm in}$. In Sect. \ref{sec|Nbody} we will estimate quantitatively the timescale for the solution to converge toward the stationary state by solving the time-dependent Fokker-Planck equation with multiplicative noise and an appropriate initial conditions gauged on $N-$body simulations.}
\begin{equation}
\mathcal{P}(M,t) = \sqrt{\frac{2}{\pi}}\, \frac{\delta_c}{\sigma^2}
\left|\frac{{\rm d}\sigma}{{\rm d}M}\right|\, e^{-\delta_c^2/2\, \sigma^2}~;
\end{equation}
the related mass function after Eq.~(\ref{eq|massfunc}) is found to be the formula originally suggested by Press \& Schechter (1974), including the fudge factor of $2$ in the normalization that in the literature is often  justified via the excursion set approach (see Bond et al. 1991).

\subsection{Multiplicative noise}

To add more complex stochastic dynamics that will be needed to fit the $N-$body mass functions and investigate universality, it is convenient to reformulate our theory in terms of the scaled variable $\nu\equiv\delta_c(t)/\sigma(M)$ and introduce a modulation of the noise term. The basic evolution equation becomes (see LD20 for details):
\begin{equation}\label{eq|basic_barrier}
\frac{\rm d}{{\rm d}t}\,\nu = -\nu\,\left|\frac{\dot\delta_c}{\delta_c}\right| +\frac{\nu}{B(\nu)}\,\left|\frac{\dot\delta_c}{\delta_c}\right|^{1/2}\, \eta(t)~,
\end{equation}
where $B(\nu)$ is a function that describe a mass/redshift dependence of the noise strength. In particular, changing variable from $\nu$ to $M$, one easily recognizes that $B(\nu)=\nu$ exactly corresponds to Eq.~(\ref{eq|basic_mass}), and in such a case the stochastic quantity $\nu$ evolves following an Ornstein–Uhlenbeck process with additive noise. If $B(\nu)\neq \nu$ instead the noise becomes multiplicative, i.e. dependent from the system's state, adding more complex dynamics. A simple yet flexible choice for the function $B(\nu)$ is
\begin{equation}\label{eq|barrier}
B(\nu) = \sqrt{q}\, \nu\, \left[1+\cfrac{\beta}{(\sqrt{q}\, \nu)^{2\,\gamma}}\right]~,
\end{equation}
in terms of three parameters $(q,\beta,\gamma)$ that will be set later on by comparison with the $N$-body mass functions. The above shape is also employed in the excursion set approach, where it renders a possible mass dependence in the critical threshold for collapse $\delta_c(M,t)\approx \delta_c(t)\, B(\nu)/\nu$. However, in our theory $B(\nu)$ is just a description for the mass/redshift dependence of the multiplicative noise, and other expressions may in principle be considered.

The mass function is plainly related to the distribution of $\nu$ values at any given cosmic time as
\begin{equation}\label{eq|massfuncnu}
N(M,t) = \cfrac{\bar \rho}{M}\, \mathcal{P}(M,t) = \cfrac{\bar \rho}{M^2}\, \left|\cfrac{{\rm d}\ln\nu}{{\rm d}\ln M}\right|\, \nu\,\mathcal{P}(\nu,t)~;
\end{equation}
the quantity $f(\nu,t)=\nu\, \mathcal{P}(\nu,t)$ is often referred in the literature as `multiplicity function'. The mass function $N(M,t)$ is said to be `universal' if the multiplicity function is only a function of $\nu$ and has no explicit time dependence, meaning that the evolution in cosmic time is solely encapsulated into the scaled variable $\nu$.

Analogously to Sect. \ref{sec|basics}, the probability distribution $P(\nu,t)$ can be derived from the  Fokker-Planck equation associated to the stochastic Eq.~(\ref{eq|basic_barrier}), which reads:
\begin{equation}\label{eq|fokker_barrier}
\partial_t\,\mathcal{P}(\nu,t)=\left|\cfrac{\dot\delta_c}{\delta_c}\right|\,
\partial_\nu\,\left\{\nu\,\mathcal{P}(\nu,t)+\cfrac{\nu}{B(\nu)}\,\partial_\nu\,
\left[\cfrac{\nu}{B(\nu)}\,\mathcal{P}(\nu,t)\right]\right\}~,
\end{equation}
supplemented by boundary conditions $\mathcal{P}(\infty,t)=0$ and $\mathcal{P}(\nu,t)=0$ for $\nu<0$, and  by an initial condition $P(\nu,t_{\rm in})=\delta_D(\nu-\nu_{\rm in})$ that will be gauged via $N-$body simulations (see Sect. \ref{sec|Nbody}). To solve Eq.~(\ref{eq|fokker_barrier}) it is useful to rescale the time in terms of a new variable $\tau\equiv -\ln(\delta_c/\delta_{c,\rm in})$, introduce the stretched space variable $x\equiv \int{\rm d}\nu\; B(\nu)/\nu$, and define a new density $\mathcal{F}\equiv (\nu/B)\, \mathcal{P}$. In this way we turn the Fokker-Planck equation in the canonical form
\begin{equation}\label{eq|fokkercan}
\partial_\tau\,\mathcal{F}(x,\tau)= \partial_x\,\left[B(x)\, \mathcal{F}(x,\tau)\right]+\partial_x^2\,\mathcal{F}(x,\tau)~.
\end{equation}

For the simple case $B(\nu)=\nu$, for which actually the rescaling $x=\nu$ and $\mathcal{F}=\mathcal{P}$ have no effect, the time-dependent solution satisfying the aforementioned initial and boundary conditions can be found analytically via a simple Fourier transform; the result reads
\begin{equation}\label{eq|solt_ou}
\mathcal{P}(\nu,t) = \cfrac{1}{\sqrt{2\pi\,(1-\xi)}}\,
\left\{\exp\left[-\cfrac{(\nu-\nu_{\rm in}\,\sqrt{\xi})^2}{2(1-\xi)}\right]+
\exp\left[-\cfrac{(\nu+\nu_{\rm in}\,\sqrt{\xi})^2}{2(1-\xi)}\right]
\right\}~,
\end{equation}
where $\xi\equiv e^{-2\,\tau}=\delta_c^2/\delta^2_{c,\rm in}$. Re-expressing $\nu=\delta_c/\sigma$ and $\nu_{\rm in}=\delta_{c,\rm in}/\sigma_{\rm in}$ and considering Eq.~(\ref{eq|massfuncnu})  this is seen to be the very same expression Eq.~(\ref{eq|sol_FPmass}). Far away from the initial conditions $\xi\ll 1$ applies, and the solution $\mathcal{P(\nu,\infty)}=\bar{\mathcal{P}}(\nu)$ converges to the stationary state $\bar{\mathcal{P}}(\nu)=\sqrt{2/\pi}\, e^{-\nu^2/2}$ yielding again the Press \& Schechter (1974) mass function.

For nonlinear $B(\nu)$ the time-dependent Fokker-Planck equation does not admit analytic treatment, and one has to rely on numerical methods. Nevertheless, when the noise strength is just a perturbation over the linear case  $B(\nu)\sim \nu$ as expected in the present context, it is possible to work out an expression that approximate the exact solution, along the lines developed in the Appendix. The result, taking into account the appropriate boundary conditions, reads
\begin{equation}\label{eq|solt}
\begin{aligned}
\mathcal{P}(\nu,t) & = \cfrac{\mathcal{A}}{2\,\sqrt{1-\xi}}\,\cfrac{B(\nu)}{\nu}\,\left\{\exp \left[-\theta\,\cfrac{\sqrt{\xi}}{1-\xi}\,\cfrac{(x-x_{\rm in})^2}{2}\right]+\exp \left[-\theta\, \cfrac{\sqrt{\xi}}{1-\xi}\,\cfrac{(x+x_{\rm in})^2}{2}\right]\right\}\times\\
\\
& \times \exp\left\{-\cfrac{1}{1+\sqrt{\xi}}\,\int^\nu{\rm d}\nu'\, \cfrac{B^2(\nu')}{\nu'}-\cfrac{\sqrt{\xi}}{1+\sqrt{\xi}}\,\int^{\nu_{\rm in}}{\rm d}\nu'\, \cfrac{B^2(\nu')}{\nu'}+\right.\\
& \\
&+ \left. \cfrac{\sqrt{\xi}}{1+\sqrt{\xi}}\,\cfrac{B^2(\nu_{\rm in})}{\theta}+\left[1-\cfrac{1}{\theta}\,\cfrac{{\rm d}\ln B(\nu)}{{\rm d}\ln\nu}\right]\,\ln(1+\sqrt{\xi})\right\}~,
\end{aligned}
\end{equation}
where $\xi\equiv e^{-2\,\theta\,\tau}=[\delta(t)/\delta(t_{\rm in})]^{2\, \theta}$, $x(\nu)\equiv \int^\nu{\rm d}\nu' B(\nu')/\nu'$, $x_{\rm in}=x(\nu_{\rm in})$, the normalization constant $\mathcal{A}$ is determined by the condition $\int_0^\infty{\rm d}\nu\, \mathcal{P}=1$, and $\theta$
is a parameter controlling the intermediate-time behavior of the solution (see Appendix for details). It is a matter of simple algebra to check that this solution reduces to Eq.~(\ref{eq|solt_ou}) in the case of $B(\nu)=\nu$.
Note that hereafter we will always refer and illustrate the results from numerically solving the time-dependent Fokker-Planck equation Eq.~(\ref{eq|fokker_barrier}), but we stress that the analytic formula Eq.~(\ref{eq|solt}) is quite effective, approximating the exact solution within $15\%$ in the relevant range of $\nu$; therefore, after gauging the noise parameters $(q,\beta,\gamma)$ and the initial condition $\nu_{\rm in}$ via $N-$body simulations (see Sect. \ref{sec|Nbody}), it can be used for quickly emulating their outputs, and for modeling/forecasting purposes in a galaxy formation context.

Far away from the initial conditions $\mathcal{P(\nu,\infty)}=\bar{\mathcal{P}}(\nu)$, the solution converges to the stationary state
\begin{equation}\label{eq|fokkersol_barrier}
\bar{\mathcal{P}}(\nu)= \mathcal{A}\,\cfrac{B(\nu)}{\nu}\,\exp\left[-\int^\nu{\rm d}\nu'\, \cfrac{B^2(\nu')}{\nu'}\right]~,
\end{equation}
which has been originally derived in LD20 by a simple integration after setting $\partial_t\mathcal{P}=0$ in Eq.~(\ref{eq|fokker_barrier}).
We point out that in our stochastic framework the mass function will become universal only far away from initial conditions, when the stationary state $\bar{\mathcal{P}}(\nu)$ above is attained. However, at a generic cosmic time the solutions of the Fokker-Planck equation for reasonable initial conditions will feature an explicit time dependence in the multiplicity function $\nu\,\mathcal{P}(\nu,t)$, that will cause the mass function to break universality.

\begin{figure}[!t]
\centering
\includegraphics[width=0.8\textwidth]{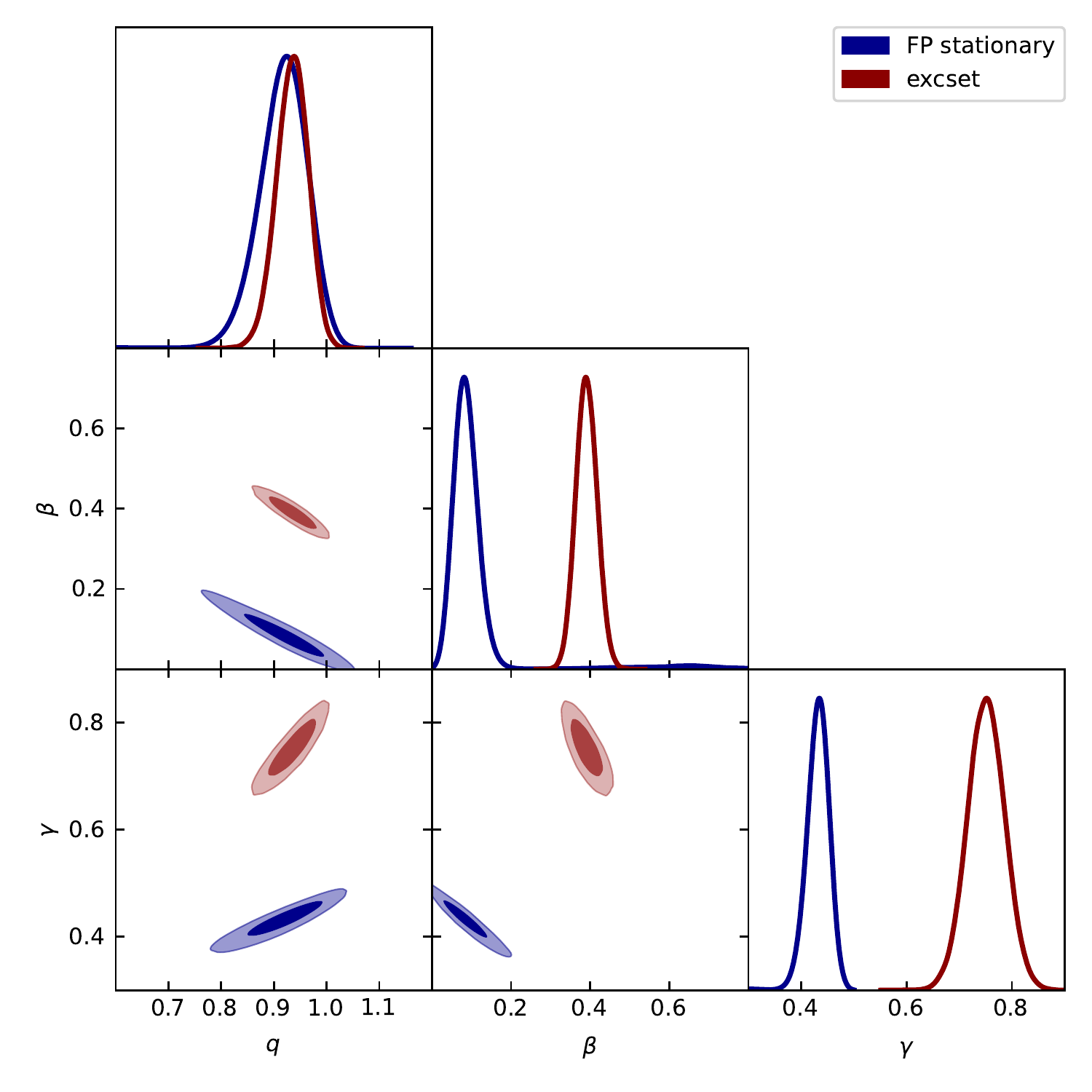}
\caption{MCMC posterior distributions for the parameters triple $(q,\beta,\gamma)$ ruling the mass/redshift dependence in Eq.~(\ref{eq|barrier}), obtained by fitting the halo multiplicity function at $z\approx 0$ by Shirasaki et al. (2021) via the stationary solutions of the Fokker-Planck equation (blue contours/lines; see Eq.~\ref{eq|fokkersol_barrier}), and via the excursion set approach (red contours/lines; see Eq.~\ref{eq|volterra}). Contours show the $68\%$ and $95\%$ confidence intervals, and the marginalized distributions are in arbitrary units (normalized to $1$ at their maximum value).}\label{fig|MCMC_steady}
\end{figure}

\section{Comparison with $N-$body mass function: non-universality as non-stationarity?}\label{sec|Nbody}

We now aim to quantitatively show the difference between stationary and non-stationary solutions, and to check whether the latter can quantitatively render the non-universality measured in the $N-$body simulations.

\begin{figure}[!t]
\centering
\includegraphics[width=\textwidth]{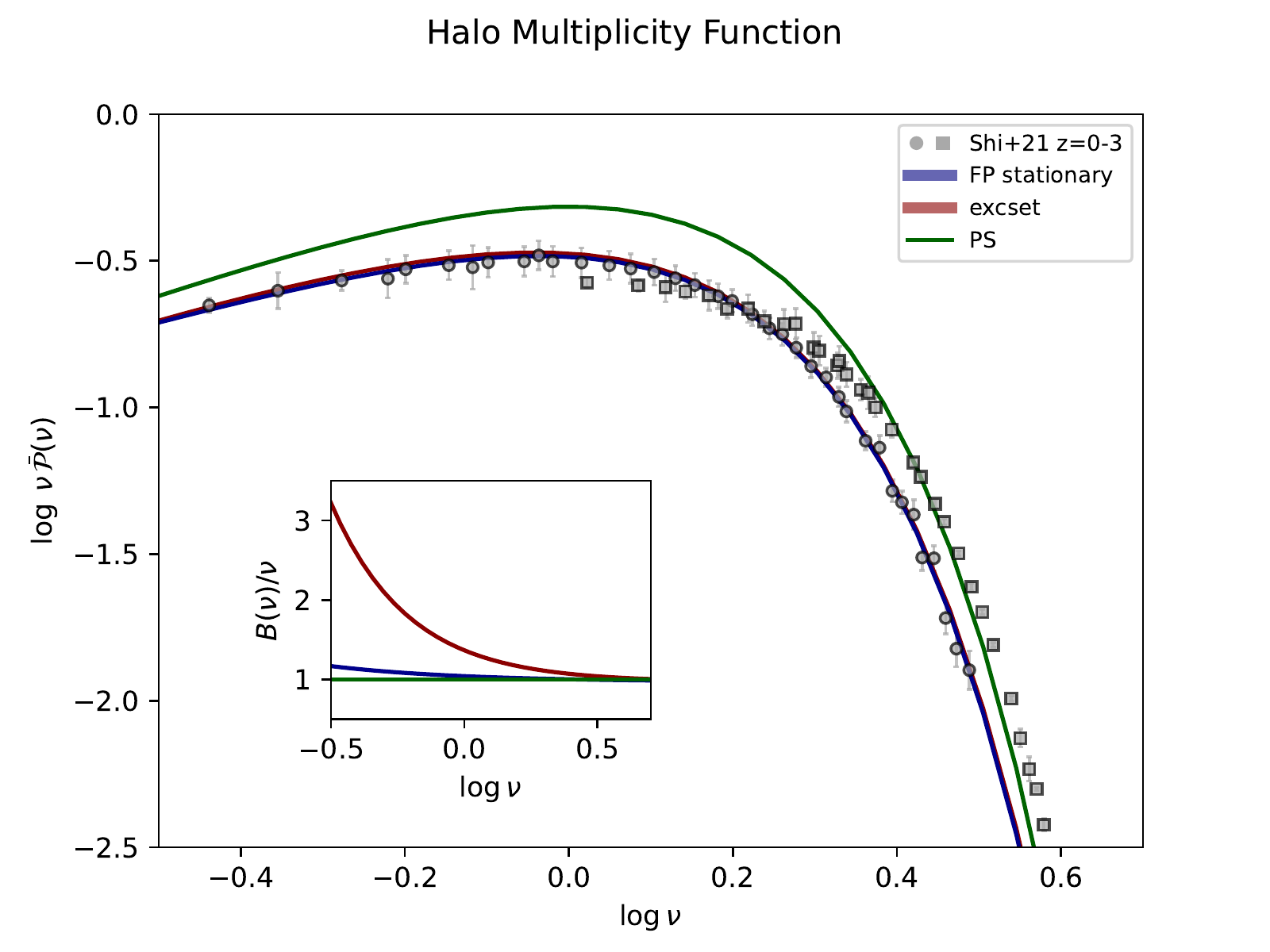}
\caption{Fits to the halo multiplicity function by Shirasaki et al. (2021) at $z\approx 0$ (circles) with the stationary solution of the Fokker-Planck equation (see Eq.~\ref{eq|fokkersol_barrier}; blue solid line), and with the excursion set approach (see Eq.~\ref{eq|volterra}; red line). The Press \& Schechter (1974) multiplicity function is also reported for reference (green lines). The non-universal behavior in the simulations is highlighted by the different shape of the multiplicity function by Shirasaki et al. (2021) at $z\approx 3$ (squares). The inset shows the deviation of the noise strength (for the stochastic theory) or of the collapse threshold (excursion set) from the constant case yielding the Press \& Schether (1974) shape.}\label{fig|MultFunc_steady}
\end{figure}

As a preliminary step, we start by fitting\footnote{For all the fits in the paper, we adopt flat priors on the parameters within the ranges $(q,\beta,\gamma)\in [0,2]$ and $\log\nu_{\rm in}\in [-2,2]$, and a standard $\chi^2$ likelihood. Then we sample the posterior distributions via a MCMC technique, by running the Python package \texttt{emcee} with $10^4$ steps and $N\times 100$ walkers where $N$ is the number of fitting parameters. Each walker is initialized with a random position uniformly-sampled from the (flat) priors. After checking the auto-correlation time we remove the first 20\% of the flattened chain to ensure burn in; the typical acceptance fractions are of order $40\%$.} the multiplicity function at $z\approx 0$ extracted from the state-of-the-art $N-$body simulations by Shirasaki et al. (2021) with the stationary solution Eq.~(\ref{eq|fokkersol_barrier}), and perform maximum likelihood estimation on the parameters $(q,\beta,\gamma)$ describing the noise strength in Eq.~(\ref{eq|barrier}). In Figs. \ref{fig|MCMC_steady} and \ref{fig|MultFunc_steady} (blue regions and lines) we illustrate the marginalized posterior distributions of the parameters (see also Table \ref{tab|MCMC_results}) and the resulting fit to the $N-$body multiplicity function at $z\approx 0$. The overall rendition of the simulated data (circles) is extremely good. In addition, the inset of Fig. \ref{fig|MultFunc_steady} shows that the fit is achieved with a minimal deviation from the simple case $B(\nu)=\nu$ (i.e., the mass and redshift dependence of the noise strength are minor) in the range of $\nu$-values probed by simulations, though the difference in the multiplicity function from the corresponding Press \& Shechter (1974) shape is appreciable.

Despite such an excellent performance, it is found that the halo mass function based on the stationary solution of the Fokker-Planck equation deviates appreciably from the $N-$body outcomes toward progressively higher redshifts; this is because, as highlighted in Fig. \ref{fig|MultFunc_steady}, the simulated multiplicity function has a slight explicit dependence on cosmic time, making its shape at $z\gtrsim 1$ considerably different from that at $z\approx 0$ (e.g., see squares in Fig. \ref{fig|MultFunc_steady} referring to $z\approx 3$). As mentioned in the previous section, in our stochastic framework a natural explanation for this additional dependence could be that the solution $\mathcal{P}(\nu,t)$ of the time-dependent Fokker-Planck Eq.~(\ref{eq|fokker_barrier}) has not yet converged to the stationary state.

To test this hypothesis, we relax the assumption of stationarity and fit the Shirasaki et al. (2021) halo multiplicity function in the redshift range $z\approx 0-3$ with the time-dependent solution of the Fokker-Planck equation. To this purpose, besides the three parameters $(q,\beta,\gamma)$ regulating the noise strength, we consider as an additional one the initial condition $\nu_{\rm in}$ needed to integrate Eq.~(\ref{eq|fokker_barrier}); the initial redshift is set at $z_{\rm in}\approx 100$ as in the simulations (anyway, we checked that the outcomes are marginally affected by the choice of $z_{\rm in}$ in the range $30-300$).

\begin{figure}[!t]
\centering
\includegraphics[width=0.8\textwidth]{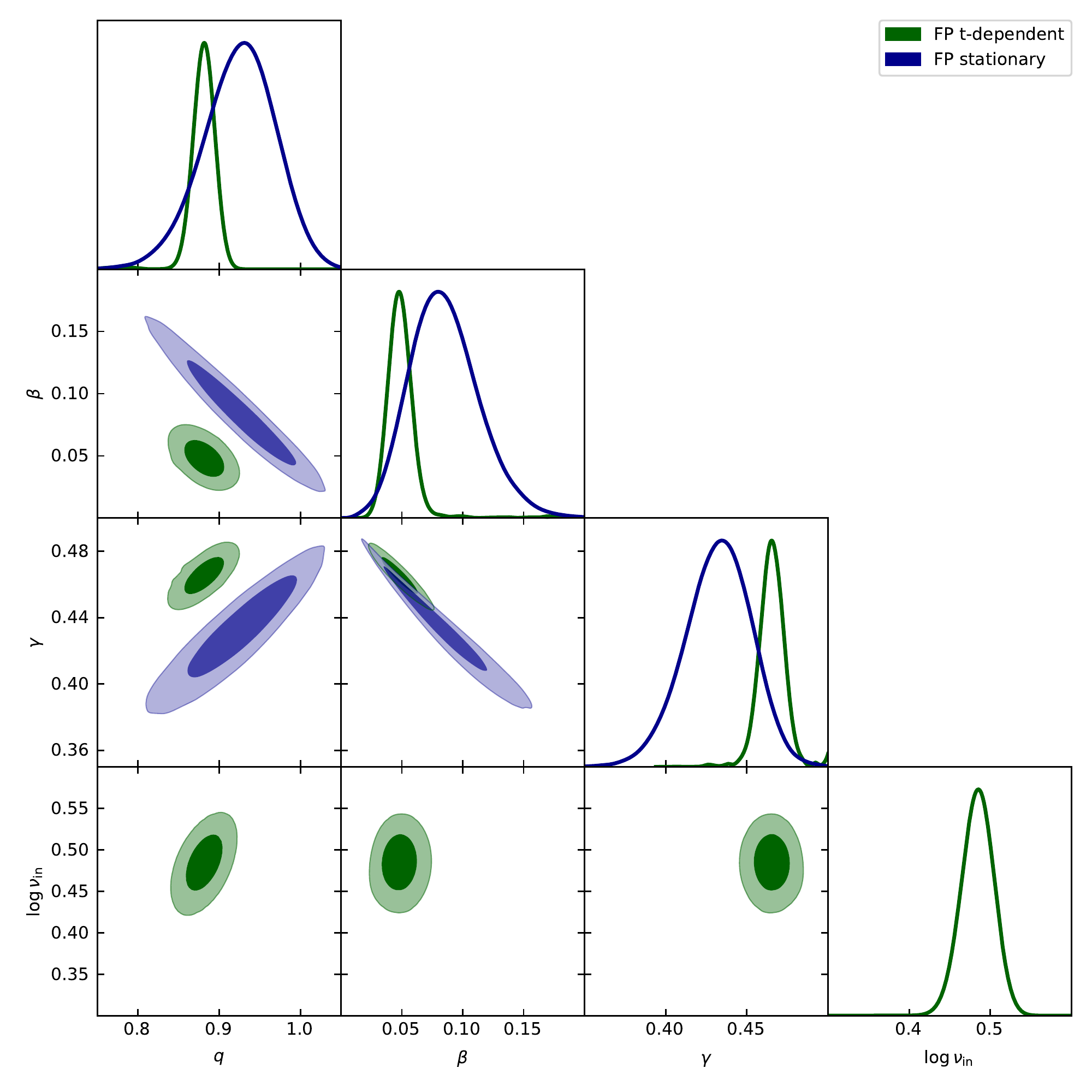}
\caption{MCMC posterior distributions for the parameters triple $(q,\beta,\gamma)$ ruling the mass/redshift dependence in Eq.~(\ref{eq|barrier}) and the initial condition $\nu_{\rm in}$ set at $z_{\rm in}\approx 100$, obtained by fitting the halo multiplicity function by Shirasaki et al. (2021) in the redshift range $z\approx 0-3$ via the time dependent solutions of the Fokker-Planck equation (green contours/lines; see Eq.~\ref{eq|solt}). For comparison, the posterior distributions on $(q,\beta,\gamma)$ for the fit via the stationary solution of the Fokker-Planck equation are reported from Fig. \ref{fig|MCMC_steady} (blue contours/lines). Contours show the $68\%$ and $95\%$ confidence intervals, and the marginalized distributions are in arbitrary units (normalized to 1 at their maximum value).}\label{fig|MCMC_tdFP}
\end{figure}

\begin{table}[h]
\centering
\begin{tabular}{|l|cccc|}
\hline
Framework & $q$ & $\beta$ & $\gamma$ & $\log \nu_{\rm in}$ \\
\hline
&&&&\\
FP - stationary  & $0.89^{+0.07}_{-0.02}$ &  $0.11^{+0.01}_{-0.06}$ & $0.42^{+0.03}_{-0.01}$ & $-$ \\
&&&&\\
Excursion set & $0.94^{+0.03}_{-0.03}$ &  $0.39^{+0.03}_{-0.03}$ & $0.75^{+0.04}_{-0.04}$ & $-$ \\
&&&&\\
FP - time dependent & $0.89^{+0.01}_{-0.02}$ &  $0.05^{+0.01}_{-0.01}$ & $0.47^{+0.01}_{-0.01}$ & $0.48^{+0.03}_{-0.01}$ \\
&&&&\\
\hline
\end{tabular}
\caption{Marginalized posterior estimates of the parameters from the MCMC analysis, for the different setups considered in the main text.}\label{tab|MCMC_results}
\end{table}

\begin{figure}[!t]
\centering
\includegraphics[width=\textwidth]{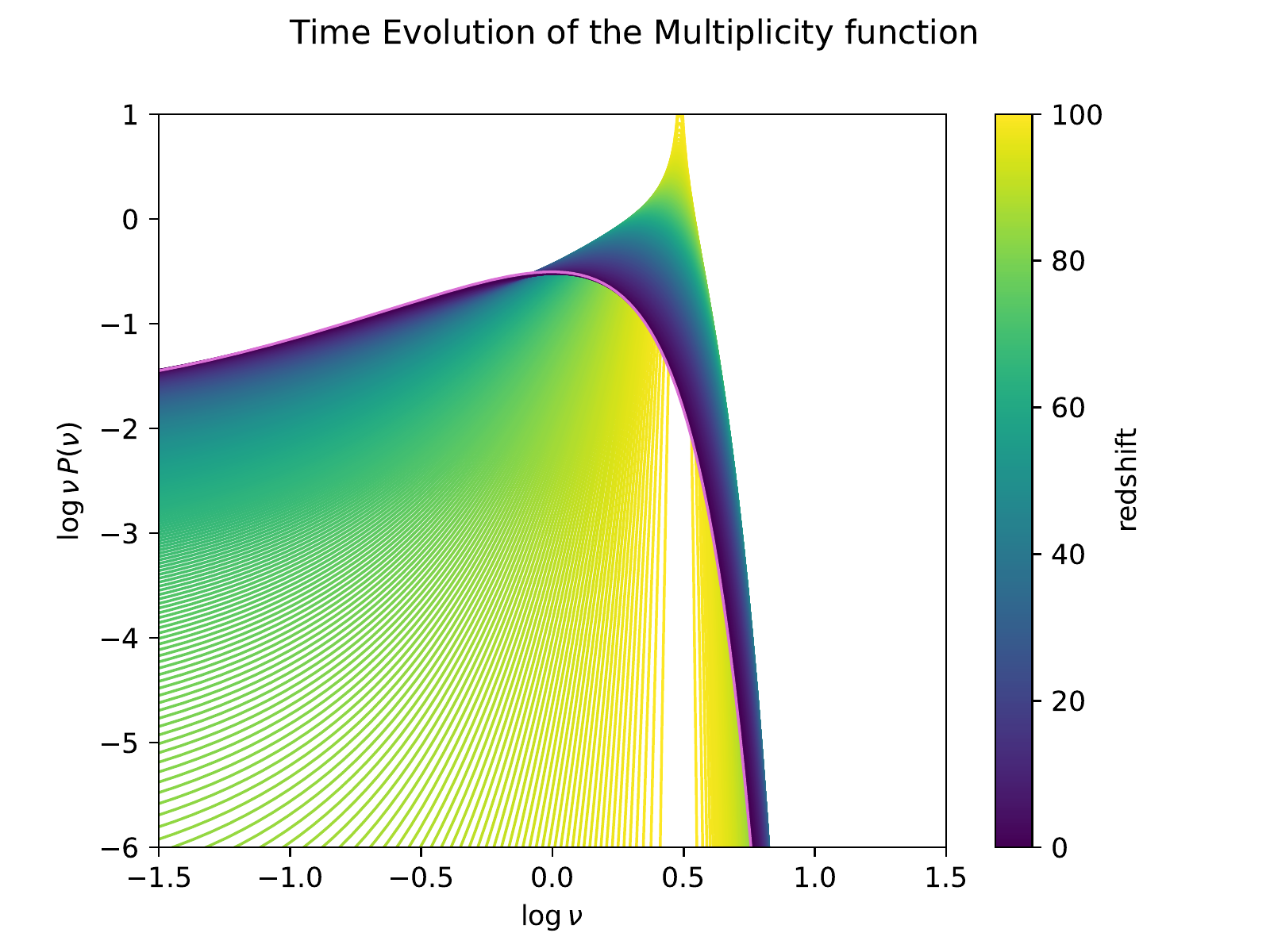}
\caption{Overall evolution of the time-dependent solutions of the Fokker-Planck equation; fitting parameters $(q,\beta,\gamma,\nu_{\rm in})$ have been set at the mean posterior values from the previous Figure (see also Table \ref{tab|MCMC_results}). Colored lines are for different redshifts from $z_{\rm in}\approx 100$ down to $z\approx 0$ as illustrated by the colorbar. The magenta line shows the corresponding stationary state.}\label{fig|MultFunc_tdFP}
\end{figure}

\begin{figure}[!t]
\centering
\includegraphics[width=\textwidth]{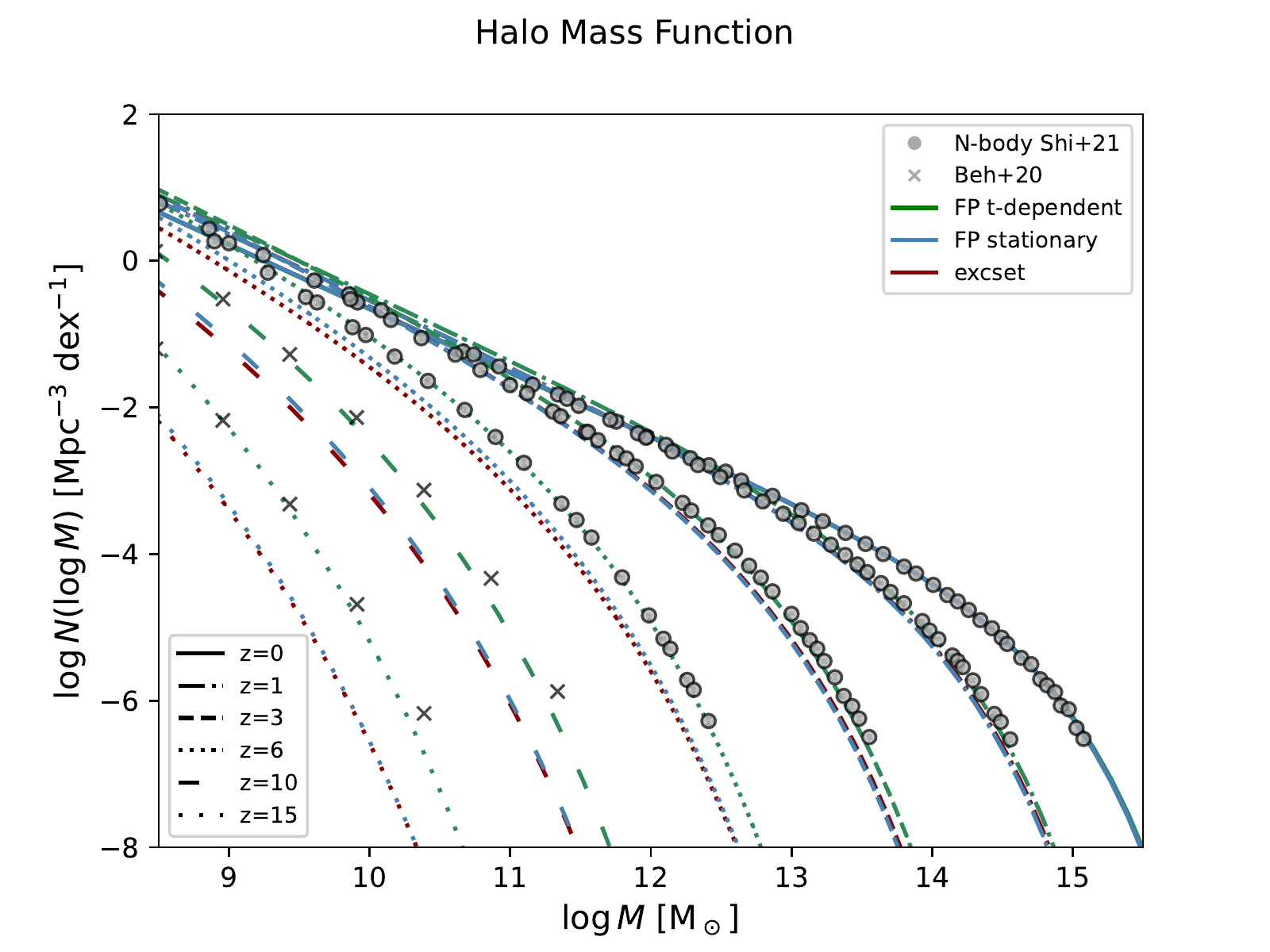}
\caption{Halo mass function at redshifts $z=0$ (solid), $1$ (dot-dashed), $3$ (dashed), $6$ (dotted), $10$ (loosely dashed), and $15$ (loosely dotted). The bestfit rendition from the time-dependent (green lines) and stationary (blue lines) solutions of the Fokker-Planck equation, and from the excursion set framework (red lines, very close to the blue ones), are compared with the outcomes from the $N-$body simulations by Shirasaki et al. (2021; grey circles) at $z\lesssim 6$.
For higher redshift $z\gtrsim 10$ we also report (sampled in $0.5$ dex mass logarithmic bins) the inference by Behroozi et al. (2013, 2020; grey crosses), which is an educated recalibration and extrapolation of the Tinker et al. (2008) simulations.}\label{fig|MassFunc_tdFP}
\end{figure}

In Fig. \ref{fig|MCMC_tdFP} we show the MCMC marginalized distributions for the fitting parameters (see also Table \ref{tab|MCMC_results}). It is seen that the values $(q,\beta,\gamma)$ are consistent within $\approx 2\sigma$ with those obtained fitting only the $z\approx 0$ multiplicity function with the stationary solution. As to the additional parameter ruling the initial condition, we find a value $\nu_{\rm in}=\delta_{c}(z_{\rm in})/\sigma(M_{\rm in})\approx 3$; given the adopted initial redshift $z_{\rm in}\approx 100$ this corresponds to a very small (earth-like) mass $M_{\rm in}\approx M_{\oplus}\approx 10^{-6}\, M_\odot$, which is pleasingly close to the free-streaming length of cold DM particles. On the one hand, this is an intriguing consistency check, since the cold nature of DM is a basic assumption of the simulations by Shirasaki et al. (2021), though the related mass resolution does not allow them to sample halos with mass below $10^5\, M_\odot$. On the other hand, specific cold DM simulations focused on the very high-redshift Universe have shown that the first structures to form in the standard cosmological framework are precisely earth-like mass halos at $z_{\rm in}\approx 50-150$ (e.g., Hoffman et al. 2001; Green et al. 2004; Diemand et al. 2005; see also review by Angulo et al. 2022), in agreement with our findings.

In Fig. \ref{fig|MultFunc_tdFP} we illustrate the evolution with redshift of the multiplicity function associated to the numerical solution of the time-dependent Fokker-Planck equation, by assuming the aforementioned bestfit parameters and initial condition. As time passes, the low-mass end (low $\nu$ values) of the multiplicity function monotonically flattens and extends towards smaller and smaller masses. Contrariwise, the behavior of the high-mass end (high $\nu$ values) first flattens and then steepens again, to produce the observed exponential suppression at low $z$.
We highlight that stationarity is marginally attained by the full time-dependent solution only toward $z\approx 0$ (see magenta line); we stress that such an evolution is an outcome of the fitting procedure and not an a-priori assumption. The implied timescale for the time-dependent solution to converge toward the stationary state amounts to many Gyrs.

Finally, in Fig. \ref{fig|MassFunc_tdFP} (see green lines) we show that the overall evolution of the $N-$body halo mass function is well captured by our time-dependent solution, much better than by the stationary state, out to high redshift. Specifically, we report the $N-$body data by Shirasaki et al. (2021) out to $z\sim 6$, which is the maximal redshift where their results can be considered robust and do not suffer of volume sampling issues.
Our non-stationary solution reproduces very well the evolution of the $N-$body mass function. This is even more remarkable on considering that the evolution of the non-stationary solutions depends mainly on one parameter, namely the initial condition $\nu_{\rm in}$, that has been calibrated by fitting the evolution of the $N-$body multiplicity function over the redshift range $z\lesssim 3$. As for higher redshifts $z\gtrsim 10$, Behroozi et al. (2013, 2020) worked out an educated recalibration and extrapolation of the $N-$body mass function from the Tinker et al. (2008) simulations; interestingly, their inference falls very close to the non-stationary solution provided by our theory.

We stress that at $z\sim 6$ the discrepancy between the stationary and non-stationary solutions amounts to a factor of a few over the whole range of relevant masses, and this increases to a factor of several or more toward higher $z\gtrsim 10$. Such a difference may have profound implications in the astrophysical and cosmological interpretation of high-redshift data from galaxy surveys, like those that will be conducted via the JWST.

\subsection{Stochastic theory vs. excursion set approach}\label{sec|stvses}

The excursion set approach has been and is still widely exploited in order to investigate the evolution of the halo mass function;  thus it is natural to compare its foundations and outcomes to those of our stochastic theory. The excursion set framework is based on the ansatz that the mass function is given by
\begin{equation}
N(M,t) = \cfrac{\bar \rho}{M^2}\, \left|\cfrac{{\rm d}\ln S}{{\rm d}\ln M}\right|\, S\, f_{\rm FC}(S)~,
\end{equation}
where $S\equiv \sigma^2$, and $f_{\rm FC}(S)$ represents the first crossing distribution of the random trajectories $\delta(S)$ across the moving barrier $\delta_c(t)\, B(\nu)/\nu=\sqrt{S}\,B(\nu)$. Note that such random walks are Markovian paths driven by white noise only when a sharp filter in Fourier space is adopted in the definition of $\sigma(M)$ via Eq.~(\ref{eq|variance}); this is the common choice in the excursion set approach, despite the technical difficulties in normalizing the filter and localizing it in real space (see Mo et al. 2010 for an educated discussion). Often in the literature the same shape in Eq.~(\ref{eq|barrier}) is adopted, and in this context it is interpreted as a mass dependent threshold due to the ellipsoidal collapse of perturbations. However, a relevant remark is that working in the abstract  $\delta-S$ space (i.e., looking for first crossing of a time-dependent barrier by a random walk) forces to consider the multiplicity and mass functions as a sequence of stationary states, the evolution in time being solely dictated by the progressive decrease of the barrier following the growth factor $\delta_c(t)\propto D^{-1}(t)$. As a consequence, the outcomes of the excursion set approach can be compared only to the stationary solutions obtained via our stochastic theory; moreover, plainly in the excursion set framework the non-universality of the mass function cannot be interpreted as non-stationarity.

The solution to the first crossing problem \footnote{Interestingly, from a historical perspective, the problem of finding the first crossing distribution to a constant barrier for a Brownian motion has been solved very early in the past century (since the first attempts by Bachelier 1900) and many studies have been  subsequently dedicated to generalize the solution for more general `moving' boundaries (e.g., Tuckwell \&  Wan 1984; Durbin 1985; for a review see Grebenkov 2015).} is implicitly given by the Volterra integral equation (see Lapi \& Danese 2013)
\begin{equation}\label{eq|volterra}
{\rm erfc}\left[\cfrac{B(S)}{\sqrt{2\, S}}\right] = \int_0^S{\rm d S'}\, f_{\rm FC}(S')\, {\rm erfc}\left[\cfrac{B(S)-B(S')}{\sqrt{2\,(S-S')}}\right]~.
\end{equation}
One of the few cases solvable analytically is the linear barrier $B_{\gamma=1}(S)=\sqrt{q}\, \delta_c+\beta\, S/\sqrt{q}\, \delta_c$ that corresponds to $\gamma=1$ in Eq.~(\ref{eq|barrier}); the solution obtained via Laplace transforms is the inverse Gaussian distribution $f_{\rm FC}(S) = \sqrt{q}\delta_c/\sqrt{2\pi\,S^3}\times e^{-B_{\gamma=1}^2(S)/2\, S}$. For generic nonlinear barrier, one must rely on numerical techniques for solving Eq.~(\ref{eq|volterra}), like the recursive method by Zhang \& Hui (2006).
In the specific case of constant barrier ($q=1$ and $\beta=0$), the Press \& Schechter (1974) mass function is recovered, which is the same outcome of the stochastic theory (in stationary conditions). This is due to a curious circumstance first pointed out by Bond et al. (1991; see their Eqs. 3.5 and 3.9): the first crossing distribution for a constant barrier can be derived
from the solution of a diffusion equation which is analogous to that of the stochastic theory (see Sect. \ref{sec|basics}).

One may ask whether the excursion set approach and our stochastic theory (for stationary solutions) produce the same results with a generic barrier like that in Eq.~(\ref{eq|barrier}). To this purpose we have exploited Eq.~(\ref{eq|volterra}) to fit the multiplicity function at $z\approx 0$ by Shirasaki et al. (2021) with the excursion set formalism, obtaining the results reported in Fig. \ref{fig|MCMC_steady} and Fig. \ref{fig|MultFunc_steady} as red lines/contours (see also Table \ref{tab|MCMC_results}). Despite the fact that the multiplicity function is fitted comparably well with respect to our stochastic theory, the excursion set approach requires the threshold for collapse to deviate substantially from the constant one (see inset), especially toward low mass halos and/or late times. On the one hand, the marginalized parameter values appreciably deviate from those $(1,0.47,0.615)$ expected from the ellipsoidal collapse of perturbations, somewhat questioning such a classic view (see Sheth \& Tormen 2002; Mo et al. 2010). On the other hand, interpreting Eq.~(\ref{eq|barrier}) as providing a mass and redshift dependent collapse threshold can cause a theoretical pitfall: the barrier required to fit simulations (see inset of \ref{fig|MultFunc_steady}) would imply a significant evolution in time at fixed mass, putting additional concerns on the implicit assumption of stationarity. Finally, as shown in Fig. \ref{fig|MassFunc_tdFP} (red lines), the excursion set framework produces intrinsically stationary solutions, that cannot explain the non-universal evolution of the mass function toward high redshift.

\section{Summary}\label{sec|summary}

In this paper we have investigated, in the framework of the stochastic theory for hierarchical clustering developed by LD20, the time-dependent solutions of the Fokker-Planck equation describing the statistics of dark matter halos. We have shown that, for reasonable initial conditions, quite long timescales of order many Gyrs are needed for such solutions to converge toward stationary states.

Although the stationary solutions can reproduce the outcomes of state-of-the-art $N-$body simulations at $z\approx 0$ to a great accuracy, requiring only a marginal dependence of the noise strength on halo mass and redshift, one needs to go beyond them to fully account for the detailed dependence on cosmic time of the simulated halo mass function. This is because the latter is found to be not strictly universal, in the sense it cannot be completely characterized in terms of a function $f(\nu)$ of a scaled variable $\nu\equiv \delta_c(t)/\sigma(M)$ encapsulating all the mass and redshift dependence via the collapse threshold $\delta_c(t)$ and the mass variance $\sigma(M)$. We have demonstrated that the time-dependent, non-stationary solutions of our stochastic theory can quantitatively reproduce, for reasonable initial conditions in the standard cosmological framework, such a non-universality and the detailed redshift dependence of the halo mass functions as measured in state-of-the-art $N-$body simulations.

Note that the issue of non-universality has been proved extremely difficult to clear in different approaches, like the standard excursion set framework. In the latter, the basic connection between the halo mass function and the first crossing distribution of a random walk hitting a mass and time-dependent barrier, requires to implicity assume stationarity; this inevitably leads to a universal behavior in the halo statistics.
Contrariwise, our stochastic theory allows for a natural and physical interpretation of the non-universality, just overwhelming any assumption of stationarity, and elucidating that to reach stationary states requires many Gyrs.

As a consequence, non-stationary and stationary solutions yield halo mass functions with appreciable differences at increasing redshift, amounting to a factor a few at $z\sim 6$ and several or more for $z\gtrsim 10$. On the one hand, such a difference could be further tested with $N-$body simulations probing an extended mass range at these substantial redshifts, which are currently challenging but will be within the reach of upcoming exa-scale supercomputing facilities. On the other hand, the non-stationarity in the halo mass function at high $z\gtrsim 10$ could be tested, though indirectly (linking luminous to dark matter is a delicate and uncertain procedure), by estimating observationally the abundance and clustering of first stars and primeval galaxies, a task that will be achieved via the analysis of JWST data (e.g., Harikane et al. 2022).

In a future perspective, the non-stationary and non-universal halo mass function from our stochastic theory could be helpful in designing, interpreting, and emulating $N-$body experiments with specific characteristics in the initial conditions, space and time resolutions, power spectra (e.g., related to nonstandard DM scenarios) and cosmological background (e.g., modified gravity theories).

\begin{acknowledgements}
We thank the referee for constructive comments and suggestions. This work has been supported by the EU H2020-MSCA-ITN-2019 Project 860744 `BiD4BESt: Big Data applications for black hole Evolution STudies' and by the PRIN MIUR 2017 prot. 20173ML3WW, `Opening the ALMA window on the cosmic evolution of gas, stars and supermassive black holes.'
\end{acknowledgements}

\begin{appendix}

\section{Approximate solution of the time-dependent Fokker-Planck equation}

In this Appendix we aim to obtain analytic expressions that approximate, \emph{uniformly} in space and time, the solutions of the time-dependent Fokker-Planck equation (see Martin et al. 2019). We focus on the case of a constant diffusion coefficient (one can take it equal to $1$, without loss of generality) and space-dependent drift so the Fokker-Planck equation to solve is
\begin{equation}\label{eq|app_FP}
\partial_\tau\,\mathcal{F}(x,\tau)= \partial_x\,\left[B(x)\, \mathcal{F}(x,\tau)\right]+\partial_x^2\,\mathcal{F}(x,\tau)~;
\end{equation}
with initial condition $\mathcal{F}(x,\tau_{\rm in})=\delta_D(x-x_{\rm in})$. This is Eq.~(\ref{eq|fokkercan}) of the main text.

First, we consider the stationary solution of Eq.~(\ref{eq|app_FP}), which can be derived by settting $\partial_\tau\,\mathcal{F}=0$; it reads
\begin{equation}\label{eq|app_FPstat}
\bar{\mathcal{F}}(x)\propto \exp\left\{-\int^x{\rm d}x'\, B(x')\right\}~.
\end{equation}
Then we introduce the function
\begin{equation}\label{eq|app_Gdef}
\mathcal{G}(x,\tau)\equiv \cfrac{\mathcal{F}(x,\tau)}{\bar{\mathcal{F}}(x)}~,
\end{equation}
that is easily seen to satisfy the backward Fokker-Planck equation
\begin{equation}\label{eq|app_Geq}
\partial_\tau\,\mathcal{G}(x,\tau)= -B(x)\,\partial_x\, \mathcal{G}(x,\tau)+\partial_x^2\,\mathcal{G}(x,\tau)~.
\end{equation}
We now define the function
\begin{equation}\label{eq|app_Hdef}
\mathcal{H}(x,\tau)\equiv -\partial_x\ln\mathcal{G}(x,\tau)~,
\end{equation}
which is found to satisfy a Burgers-like equation
\begin{equation}\label{eq|app_Heq}
\partial_\tau\,\mathcal{H}(x,\tau)= \partial_x\,\left[-B(x)\, \mathcal{H}(x,\tau)-\mathcal{H}^2(x,\tau)+\partial_x\,\mathcal{H}(x,\tau)\right]~.
\end{equation}

For the simple case $B(x)=\theta\,x$ it is easily verified by substitution in Eq.~(\ref{eq|app_Heq}) that the solution is
\begin{equation}\label{eq|app_Hou}
\mathcal{H}(x,\tau)= \theta\,\cfrac{\xi\, x-\sqrt{\xi}\, x_{\rm in}}{1-\xi} = \theta\,\cfrac{\sqrt{\xi}}{1-\xi}\,(x-x_{\rm in})-\cfrac{\sqrt{\xi}}{1+\sqrt{\xi}}\, \theta\,x~,
\end{equation}
with $\xi\equiv e^{-2\,\theta\,\tau}$. The first term in the last expression is dominant for $\tau\rightarrow 0$ (or $\xi\rightarrow 1$), yielding $\mathcal{H}(x,\tau)\simeq (x-x_{\rm in})/2\,\tau$ which for the original density $\mathcal{F}$ corresponds to the initial condition of a Dirac delta centered in $x_{\rm in}$. This term must always be present also for general $B(x)$; hence we are led to make the ansatz
\begin{equation}\label{eq|app_Hsol}
\mathcal{H}(x,\tau)\simeq \theta\,\cfrac{\sqrt{\xi}}{1-\xi}\,(x-x_{\rm in})-\cfrac{\sqrt{\xi}}{1+\sqrt{\xi}}\, B(x)~.
\end{equation}
Remarkably, when inserting this expression into Eq.~(\ref{eq|app_Heq}) and  Laurent expanding around $\tau\rightarrow 0$, the l.h.s. and r.h.s. agree at orders $\tau^{-2}$ and $\tau^{-1}$, so that the ansatz is correct at order $\tau^0$ in the short-time limit (in principle one could improve the approximation by adding a Taylor series around $\tau=0$ on the r.h.s. of Eq.~(\ref{eq|app_Hsol}), but we will not pursue this here). However, note that now the expansion of Eq.~(\ref{eq|app_Hsol}) reads $\mathcal{H}(x,\tau)\simeq (x-x_{\rm in})/2\, \tau + B(x)/2$ so it is actually independent of $\theta$; this means that $\theta$ is arbitrary in some sense and must be inferred by some other means. We will come back to this issue below.

We now determine $\mathcal{G}$, up to a time-dependent normalization constant, by integrating Eq.~(\ref{eq|app_Hdef}), to yield
\begin{equation}\label{eq|app_Gform}
\mathcal{G}(x,\tau)\simeq \mathcal{N}(\tau)\, \exp\left\{-\int_{x_{\rm in}}^x{\rm d}x'\,\mathcal{H}(x',\tau)\right\}~.
\end{equation}
Using the expression for $\mathcal{H}(x,\tau)$ from Eq.~(\ref{eq|app_Hsol}) we get
\begin{equation}\label{eq|app_Gsol}
\mathcal{G}(x,\tau)\simeq \mathcal{N}(\tau)\, \exp\left\{-\theta\,\cfrac{\sqrt{\xi}}{1-\xi}\,\cfrac{(x-x_{\rm in})^2}{2}+\cfrac{\sqrt{\xi}}{1+\sqrt{\xi}}\, \int^x_{x_{\rm in}}{\rm d}x'\, B(x')\right\}~.
\end{equation}

The dependence on the starting point $x_{\rm in}$ in the normalization constant can be elicited by inserting Eq.~(\ref{eq|app_Gform}) in Eq.(\ref{eq|app_Geq}) and obtaining the first-order differential equation
\begin{equation}
\cfrac{\dot{\mathcal{N}}}{\mathcal{N}} =  \int_{x_{\rm in}}^x{d}x'\,  \partial_\tau\, \mathcal{H}(x',\tau)+ B(x)\,\mathcal{H}(x,\tau)-\partial_x\,\mathcal{H}(x,\tau)+\mathcal{H}^2(x,\tau)~.
\end{equation}
The r.h.s. appears also to depend on $x$ but actually do not; in fact, using Eq.~(\ref{eq|app_Heq}) and considering that by definition $\mathcal{N}(\tau)\rightarrow 1$ for $\tau\rightarrow \infty$ one finds
\begin{equation}
\mathcal{N}(\tau) = \exp\left\{\int_\tau^\infty{\rm d}\tau'\, \left[-B(x)\,\mathcal{H}(x,\tau')+\partial_x\,\mathcal{H}(x,\tau')-\mathcal{H}^2(x,\tau')\right]|_{x=x_{\rm in}}\right\}~.
\end{equation}

Exploiting now the explicit form of $\mathcal{H}(x,\tau)$ given by Eq.~(\ref{eq|app_Hsol}), after some algebra, one finds
\begin{equation}
\mathcal{N}(\tau) = \cfrac{1}{\sqrt{1-\xi}}\, \exp\left\{\cfrac{\sqrt{\xi}}{1+\sqrt{\xi}}\,\cfrac{B^2(x_{\rm in})}{\theta}+\left[1-\cfrac{B'(x_{\rm in})}{\theta}\right]\,\ln(1+\sqrt{\xi})\right\}~.
\end{equation}

Using the above normalization in Eq.~(\ref{eq|app_Gsol}), and combining Eqs. (\ref{eq|app_FPstat}) and (\ref{eq|app_Gdef}) we obtain
\begin{equation}\label{eq|app_Fsol}
\begin{aligned}
\mathcal{F}(x,\tau) \propto  \cfrac{1}{\sqrt{1-\xi}}\,\exp & \left\{- \theta\,\cfrac{\sqrt{\xi}}{1-\xi}\,\cfrac{(x-x_{\rm in})^2}{2}-\cfrac{1}{1+\sqrt{\xi}}\,\int^x{\rm d}x'\, B(x')-\cfrac{\sqrt{\xi}}{1+\sqrt{\xi}}\,\int^{x_{\rm in}}{\rm d}x'\, B(x')\right.\\
& \\
&+ \left. \cfrac{\sqrt{\xi}}{1+\sqrt{\xi}}\,\cfrac{B^2(x_{\rm in})}{\theta}+\left[1-\cfrac{B'(x_{\rm in})}{\theta}\right]\,\ln(1+\sqrt{\xi})\right\}~.\\
\end{aligned}
\end{equation}

We are now left with the problem of inferring the parameter $\theta$, that controls the intermediate time behavior of the solution. One can argue to choose $\theta$ so as to minimize the error in the expansion underlying Eq.~(\ref{eq|app_Hsol}), that occurs for $B'(x)\simeq \theta$; this suggests to take
\begin{equation}\label{eq|app_theta}
\theta \simeq \int{\rm d}x\,{\bar{\mathcal{F}}}(x)\, B'(x)~,
\end{equation}
in terms of the steady-state solution ${\bar{\mathcal{F}}}(x)$.
It is straightforward to verify that for $B(x)=x$ the above yields $\theta=1$ and Eq.~(\ref{eq|app_Fsol}) collapses into the form
\begin{equation}
\mathcal{F}(x,\tau) \propto  \cfrac{1}{\sqrt{1-\xi}}\,\exp \left\{-\cfrac{(x-\sqrt{\xi}\,x_{\rm in})^2}{2\,(1-\xi)}\right\}~;
\end{equation}
this is in fact the classic solution for the Fokker-Planck equation with constant diffusion coefficient and linear drift, that corresponds to an Ornstein–Uhlenbeck stochastic process with additive noise.

\end{appendix}

\end{document}